\documentclass[prb,reprint]{revtex4-1}

\usepackage{graphicx}
\usepackage{bm}

\begin{document}

\title{Statistics of heat exchange between two resistors}
\author{D.S. Golubev and J.P. Pekola}
\address{Low Temperature Laboratory, 
Department of Applied Physics, Aalto University School of Science, P.O. Box 13500, 00076 AALTO, Finland}
\begin{abstract}
We study energy flow between two resistors coupled by an arbitrary 
linear and lossless electric circuit. We show that the fluctuations of energy
transferred between the resistors are determined
by random scattering of photons on an effective barrier with frequency
dependent transmission probability $\tau(\omega)$. We express the latter 
in terms of the circuit parameters. Our results are valid 
in both quantum and classical regimes and for non-equilibrium electron 
distribution functions in the resistors. Our theory is in good agreement with recent
experiment performed in the classical regime.
\end{abstract}

\maketitle

\section{Introduction}

The problem of energy exchange between two resistors  
has been first analyzed by Nyquist  \cite{Nyquist} on the way  
towards his famous formula for the current noise of a resistor,  
\begin{eqnarray}
S_I = 4k_BT/R.
\label{Nyquist}
\end{eqnarray}
Here $S_I$  is the spectral density of noise at low frequencies $|\omega|\ll k_BT/\hbar$,
$k_B$ is the Boltzmann constant, $T$ is the temperature and $R$ is the resistance. 
Equation (\ref{Nyquist}) has been confirmed by Johnson \cite{Johnson} and by numerous 
subsequent experiments. For a long time afterwards transport of heat in electric 
circuits has been considered well understood. 
Recently, however, it has attracted renewed attention due to advances both in theory
and in technology. On the theoretical side, the discovery of the fluctuation 
theorem \cite{BK,Evans,Crooks,Campisi} has triggered the interest
in the statistics of heat transport. Statistics of effective electron temperature fluctuations
in small metallic grains is also under discussion \cite{Nazarov1,Nazarov2}.
The experiments have recently
advanced in two directions. First, quantum transport of heat between two 
resistors coupled by superconducting wires and separated by up to 50 $\mu$m distance 
has been demonstrated at sub-kelvin temperatures \cite{Meshke,Timofeev}.
Second, utilizing low noise amplifiers Ciliberto {\it et al.} have recently 
measured the full statistical distribution of heat transferred between two resistors
kept at temperatures 88 K and 296 K respectively \cite{Ciliberto1,Ciliberto2}.
They have verified the validity of the fluctuation theorem 
and worked out a theoretical model based on 
Nyquist's formula (\ref{Nyquist}).

\begin{figure}
\includegraphics[width=8.5cm]{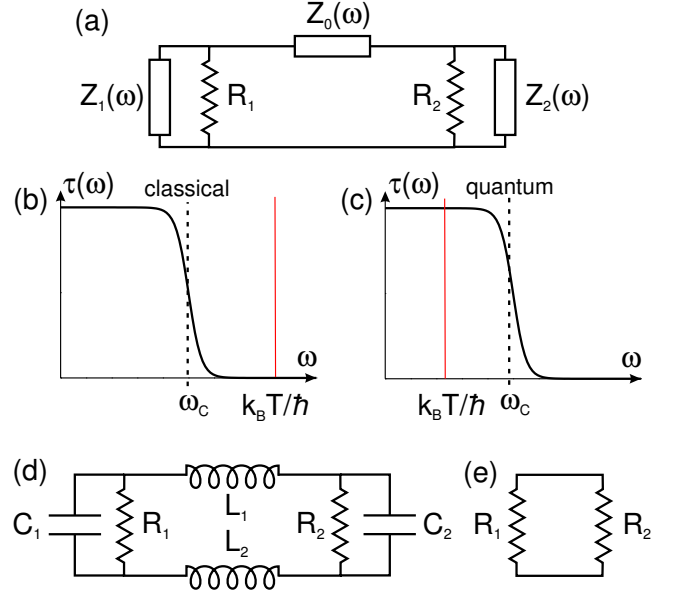}
\caption{Two resistors connected in a linear circuit: 
(a) general case -- resistors are connected by
an arbitrary reactive element with the impedance $Z_0(\omega)$ and shunted by the reactive impedances $Z_1(\omega),Z_2(\omega)$;
(b) classical regime $T_1, T_2\gtrsim \omega_c$; 
(c) quantum regime $T_1,T_2\lesssim \omega_c$; 
(d) realistic model, with stray capacitances $C_1,C_2$ and wire inductances $L_1,L_2$;
(e) resistors directly coupled by two ideal zero resistance wires.}
\end{figure}

Motivated by these developments, in this letter we propose a theory of full counting
statistics of photon mediated heat exchange between two metallic resistors valid both at high and 
at low temperatures, where the classical formula
for the noise (\ref{Nyquist}) can no longer be used. 
We consider two resistors, $R_1$ and $R_2$ shunted by impedances $Z_1(\omega)$ and $Z_2(\omega)$, 
and coupled by a linear element (e.g. transmission line, capacitor, etc.) 
having the impedance $Z_0(\omega)$  (see Fig. 1a).
The impedances $Z_j(\omega)$, ($j=0,1,2$) are purely reactive
and do not generate noise.  The average
photonic heat current flowing from the resistor 1  
to the resistor 2 reads
\begin{eqnarray}
J_Q=\int_0^\infty \frac{d\omega}{2\pi}\,\omega\tau(\omega)\big[n_1(\omega)-n_2(\omega)\big],
\label{P}
\end{eqnarray}
where $\tau(\omega)$ is the effective transmission, which we will specify later, $n_j(\omega)$
are photon distribution functions (here and below we put $k_B=\hbar =1$). 
Typically $\tau(\omega)$ drops at certain cutoff frequency $\omega_c$. Assuming that $n_1(\omega),n_2(\omega)$
have equilibrium Bose form with the temperatures $T_1$ and $T_2$, one finds that at high temperatures,
$T_1,T_2\gtrsim \omega_c$ (Fig. 1b), $J_Q\approx \tau(0)\omega_c(T_1-T_2)$ in agreement
with experimental findings of Refs. \cite{Ciliberto1,Ciliberto2}. In this classical regime 
Nyquist's formula (\ref{Nyquist}) may be used to derive the heat current. 
In this letter we will be mostly interested in the opposite, quantum, limit $T_1,T_2\lesssim \omega_c$
(Fig. 1c), which is relevant for typical low temperature experiments \cite{Meshke,Timofeev}. 
Indeed, the cutoff frequency may be estimated as 
$\omega_c\sim \min\{{1/R_jC_j, R_j/L_j}\}$, where $C_j\sim \epsilon\epsilon_0 l$ are
stray capacitances, $L_j\sim \mu_0 l$ are inductances of the wires (Fig. 1d), 
$\epsilon_0$ and $\mu_0$ are vacuum permittivity and permeability, $\epsilon$ is the dielectric constant,
and $l$ is the characteristic 
size of the sample. For the parameters of the low temperature experiments  \cite{Meshke,Timofeev}, namely 
$T\sim 100$ mK, $R\sim 1$ k$\Omega$ and $l\sim 10$ $\mu$m, one
finds $T/\omega_c\sim 10^{-3}\ll 1$. Thus the circuit is in the quantum regime. 
In contrast, for the experiments by Ciliberto {\it et al} \cite{Ciliberto1,Ciliberto2}
with $T\sim 100$ K, $R=10$ M$\Omega$ and $l\sim 1$ cm one finds $T/\omega_c\sim 10^{10}\gg 1$, 
which corresponds to strongly classical regime. 

\section{Model}

Our goal is to find the distribution of the energy $Q$ transferred from the resistor 1 to the resistor 2 during the time $t$, 
which we denote as $P(t,Q)$. 
It is more convenient to work with the cumulant generating function (CGF), $F(t,\lambda)$,
which depends on the counting field $\lambda$ and defined as
\begin{eqnarray}
e^{F(t,\lambda)}=\int dQ\, e^{i\lambda Q} P(t,Q).
\label{PQ}
\end{eqnarray}
We describe the system by a Hamiltonian
\begin{eqnarray}
\hat H=\hat H_0  + \hat H_{\rm em} + \hat H_{\rm int},
\end{eqnarray}
where $\hat H_0=\sum_{k\sigma} \epsilon_k \hat a^\dagger_{k\sigma} \hat a_{k\sigma}$ is the Hamiltonian
of non-interacting electrons moving in the combined potential of ion lattice and impurities, $\hat a_{k\sigma}$
is an annihilation operator of an electron in the eigenstate $|\psi_{k\sigma}\rangle$ ($\sigma$ is the spin index) and $\epsilon_k$
is the corresponding eigen-energy;
$
\hat H_{\rm em}=\int d^3{\bm r}(\hat{\bm E}^2+\hat{\bm H}^2)/8\pi
$ 
is the Hamiltonian of electro-magnetic field;
$\hat{\bm E}$ and $\hat{\bm H}$ are the operators of the electric and magnetic fields respectively;
$
\hat H_{\rm int}=-\sum_{kn,\sigma} e\hat V_{kn}\hat a^\dagger_{k\sigma}\hat a_{n\sigma}
$
is the interaction Hamiltonian; and $\hat V_{kn}=\langle \psi_k|\hat V(\bm{r})|\psi_n\rangle$ 
are the matrix elements of the electric potential operator between
two eigenfunctions of the non-interacting electron Hamiltonian $\hat H_0$.
The Hamiltonian $\hat H_0$ describes the two resistors, the
wires connecting them and the leads attached to them if they present.

An important point is the definition of the transferred energy $Q$. 
Here we have in mind the detection scheme based on normal metal - 
superconductor tunnel junctions attached to the resistors \cite{Meshke,Timofeev}. 
Such a junction allows one to measure the effective temperature of a resistor or, more generally, the distribution
function, $f(E,{\bm r})$, of electrons in it \cite{Pothier}. 
The latter can be converted into the total electron energy of the resistor $j$ ($j=1,2$) 
as ${\cal E}_j=2\int_{\Omega_j} d^3{\bm r}\int dE\, E\nu_j(E)f(E,{\bm r})$ (here
$\Omega_j$ in the volume of the resistor $j$ and $\nu_j(E)$ is the density of 
states). Within this approach it is natural to define the transferred energy 
as the drop in the electronic energy of the resistor 1 during the time $t$,    
$Q=-{\cal E}_1(t)+{\cal E}_1(0)$. 
The corresponding quantum expression for the CGF reads \cite{Campisi}:
\begin{eqnarray}
e^{F(t,\lambda)} = \,{\rm tr}\,\left[ e^{-i\lambda \hat H_{1}} e^{-i\hat Ht} e^{i\lambda \hat H_{1}} \hat\rho_0 e^{i\hat Ht} \right],
\label{F}
\end{eqnarray}
where $\hat\rho_0$ is the initial density matrix and
$\hat H_1$ is the free electron part of the Hamiltonian of the resistor 1.

The trace in Eq. (\ref{F}) can be expressed as a path integral
over the fluctuating potentials $V^{F},V^B,{\bm A}^{F},{\bm A}^{B}$ defined on the forward ($F$) and
backward ($B$) branches of the Keldysh contour, and over the Grassman fields $a^F_{k\sigma},a^{F*}_{k\sigma},a_{k\sigma}^B,a^{B*}_{k\sigma}$
describing electrons.
Performing the Gaussian integral over the latter, we get
\begin{eqnarray}
e^{F} = \int{\cal D}V^{F,B}{\cal D}{\bm A}^{F,B} \,e^{iS^\lambda[V^{F,B},{\bm A}^{F,B}]},
\label{F2}
\end{eqnarray}
where the effective action $iS^\lambda[V^{F,B},{\bm A}^{F,B}]$ is the sum
of the electronic and electromagnetic contributions,
\begin{eqnarray}
iS^\lambda &=& iS_{\rm el}^\lambda +iS_{\rm em},
\label{S}
\\
iS_{\rm el}^\lambda &=& 2\ln[\det (\check G^{-1}[V^F,V^B])],
\label{Sel}
\\
iS_{\rm em}&=& i\int_0^t dt'\int d^3{\bm r}\frac{E_F^2-E_B^2-H_F^2+H_B^2}{8\pi}.
\label{Sem}
\end{eqnarray}
Here we introduced the inverse Keldysh Green function of electrons
$\check G^{-1}_{kn} = \check G^{-1}_{0,kn} + \delta\check G^{-1}_{kn}$, where
\begin{eqnarray}
\check G^{-1}_{0,kn} &=& \delta_{kn}\left(\begin{array}{cc}  
i\partial_t-\epsilon_k &  0 \\
0 & -i\partial_t+\epsilon_k \end{array}\right),
\nonumber\\
\delta\check G^{-1}_{kn} &=& \left(\begin{array}{cc}  
e V^F_{kn}e^{-i\lambda_k\epsilon_k+i\lambda_n\epsilon_n} & 0 \\
 0 &   - e V^B_{kn}\end{array}\right).
\end{eqnarray}
At this stage we 
retain the information about occupation numbers of all energy levels
keeping the dependence of the counting filed $\lambda_k$ on the level index $k$. 
Below we will only consider linear circuits free of highly resistive junctions or quantum dots
in the Coulomb blockade regime. Then one can expand the action (\ref{Sel}) 
to the second order in  $V^F,V^B$,  
\begin{eqnarray}
iS_{\rm el}^\lambda \to  2\ln[\det\check G^{-1}_0]
+ {\rm tr}\left[2 \check G_0\delta\check G^{-1}-\left(\check G_0\delta\check G^{-1}\right)^2\right].
\label{expansion}
\end{eqnarray}  
This expression contains the Green function of non-interacting electrons, $\check G_0$. It is defined as 
\begin{eqnarray}
&&\check G_{0,kn}(t_1,t_2)=-i\delta_{kn}e^{-i\epsilon_k(t_1-t_2)}
\nonumber\\ &&\times\,
\left(\begin{array}{cc} 
\theta_{12}(1-f_k)-\theta_{21}f_k & -f_k \\ 1-f_k & -\theta_{12}f_k +\theta_{21}(1-f_k)
\end{array}\right),
\nonumber
\end{eqnarray}
where $\theta_{ij}=\theta(t_i-t_j)$ are Heaviside functions and  
$f_k=\langle\hat a^\dagger_{k\sigma}\hat a_{k\sigma}\rangle$ are the occupation numbers of
the energy levels. The first term in the expansion (\ref{expansion}) does not depend on $\lambda_k$
and may be omitted. The second term, ${\rm tr}\left[2 \check G_0\delta\check G^{-1}\right]$, 
is canceled by a similar contribution coming from positively charged ion background.
Thus, only the last term of Eq. (\ref{expansion}) matters. We transform it to the from
\begin{eqnarray}
iS_{\rm el}^\lambda &=&  e^2\int_0^t dt'dt'' \sum_{kn}\sum_{\alpha,\beta=\pm} e^{-i(\epsilon_k-\epsilon_n)(t'-t'')}
\nonumber\\ &&\times\,
\chi_{kn}^{\alpha\beta} \, V_{nk}^\beta(t')V_{kn}^\alpha(t'').
\label{action}
\end{eqnarray}
Here we have introduced the potentials $V^+=(V^F+V^B)/2$ and $V^-=V^F-V^B$,  
as well as dimensionless combinations $\chi_{kn}^{\alpha\beta}$ containing electronic distribution functions $f_k$
and counting fields $\lambda_k$: 
\begin{eqnarray}
\chi_{kn}^{++} &=& f_k(1-f_n)\left(e^{-i\lambda_k\epsilon_k+i\lambda_n\epsilon_n}-1\right)
\nonumber\\ &&
+\, (1-f_k)f_n\left(e^{i\lambda_k\epsilon_k-i\lambda_n\epsilon_n}-1\right),
\nonumber\\
\chi_{kn}^{+-} &=& (\theta_{12}-\theta_{21})(f_k-f_n)+f_k(1-f_n)e^{-i\lambda_k\epsilon_k+i\lambda_n\epsilon_n}
\nonumber\\ &&
-\, (1-f_k)f_ne^{i\lambda_k\epsilon_k-i\lambda_n\epsilon_n},
\nonumber\\
\chi_{kn}^{-+} &=& 0,
\nonumber\\
\chi_{kn}^{--} &=& -f_k(1-f_n)\left(e^{-i\lambda_k\epsilon_k+i\lambda_n\epsilon_n}+1\right)/4
\nonumber\\ &&
-\,(1-f_k)f_n\left(e^{i\lambda_k\epsilon_k-i\lambda_n\epsilon_n}+1\right)/4.
\end{eqnarray}

Next we perform disorder averaging of the matrix elements $V_{kn}^\alpha$ in Eq. (\ref{action}) 
inside the metallic parts of the system ignoring weak localization and
utilizing the rule of averaging for the product of electronic wave functions \cite{ABG}
\begin{eqnarray}
\sum_{kn}\left\langle \psi_k^*({\bm r}_2)\psi_n({\bm r}_2)\psi_k({\bm r}_1)\psi_n^*({\bm r}_1) \delta(E_1-\epsilon_k)\delta(E_2-\epsilon_n)\right\rangle
\nonumber\\
=({\nu}/{\pi})\,{\rm Re}\,{\cal D}(E_1-E_2,{\bm r}_1,{\bm r}_2).\hspace{0.5cm}
\end{eqnarray}
Here $\nu$ is the density of states and ${\cal D}(E,{\bm r}_1,{\bm r}_2)$ is 
the solution of the diffusion equation
$
(-iE-D({\bm r})\nabla^2){\cal D} = \delta({\bm r}_1-{\bm r}_2),
$
where $D({\bm r})$ is the diffusion constant. In good metals with local current-field relation,  
${\bm j}=\sigma({\bm r}){\bm E}$, where $\sigma({\bm r})=2e^2\nu_0({\bm r})D({\bm r})$ is the conductivity,
one can approximate ${\rm Re}\,{\cal D}(E,{\bm r}_1,{\bm r}_2)\to -D({\bm r})\nabla^2/E^2$, 
and the action (\ref{action}) acquires the form
\begin{eqnarray}
iS_{\rm el}^\lambda &=&  -\int_0^t dt'dt''\int d^3{\bm r}\sigma({\bm r})
\int\frac{d\omega}{2\pi} \frac{e^{-i\omega(t'-t'')}}{\omega}
\nonumber\\ &&\times\,
\sum_{\alpha,\beta=\pm} \eta^{\alpha\beta}_{\omega,{\bm r}} \nabla V^\beta(t',{\bm r})\nabla V^\alpha(t'',{\bm r}). 
\label{action1}
\end{eqnarray}
Here 
\begin{eqnarray}
\eta^{++}_{\omega,{\bm r}}&=& -n_{\omega,{\bm r}}(e^{i\lambda_{\bm r}\omega}-1)
-(n_{\omega,{\bm r}}+1)(e^{-i\lambda_{\bm r}\omega}-1),
\nonumber\\ 
\eta^{+-}_{\omega,{\bm r}}&=&1-
n_{\omega,{\bm r}}(e^{i\lambda_{\bm r}\omega}-1)
+(n_{\omega,{\bm r}}+1)(e^{-i\lambda_{\bm r}\omega}-1),
\nonumber\\ 
\eta^{-+}_{\omega,{\bm r}} &=& 0,
\nonumber\\
\eta^{--}_{\omega,{\bm r}}&=&\frac{n_{\omega,{\bm r}}(e^{i\lambda_{\bm r}\omega}+1)+(n_{\omega,{\bm r}}+1)(e^{-i\lambda_{\bm r}\omega}+1)}{4},
\label{eta}
\end{eqnarray}
and $n_{\omega,{\bm r}}$ is the effective photon distribution function,
\begin{eqnarray}
n_{\omega,{\bm r}}=\frac{1}{\omega}\int dE f\left(E+\frac{\omega}{2},{\bm r}\right)\left[1-f\left(E-\frac{\omega}{2},{\bm r}\right)\right].
\label{n_photon}
\end{eqnarray}
It satisfies $n_{-\omega,{\bm r}}=-1-n_{\omega,{\bm r}}$ and in local equilibrium, i.e. for 
momentum isotropic electron distribution function of the form $f(E,{\bm r}) =1/(e^{E/T({\bm r})}+1)$,
where $T(\bm{r})$ is the local electron temperature, 
it reduces to Bose function $1/(e^{\omega/T({\bm r})}-1)$. 
However, $n_{\omega,{\bm r}}$ may deviate from simple Bose form if the electron distribution
function is driven out of equilibrium by, for example, bias voltage applied to a resistor \cite{Pothier}.    
In Eq. (\ref{action1}) we have also assumed that the counting field $\lambda_{\bm r}$ is the same for all 
energy levels with wave functions localized in the vicinity of the point ${\bm r}$ and that it slowly varies in space at distances
exceeding the spatial extension of these wave functions.

We are now in position to write down the action of two coupled resistors
depicted in Fig. 1a. We put $\lambda({\bm r})=\lambda_j$, $\sigma({\bm r})=\sigma_j$ $(j=1,2)$ inside each
resistor. Considering low frequency modes, we also put $\nabla V({\bm r})=V_j/L_j$, where $V_j$ is the
instantaneous voltage drop across the $j-$th resistor, and $L_j$ is its length. We also define the 
resistances $R_j=L_j/\sigma_j{\cal A}_j$, where ${\cal A}_j$ are the cross-sectional areas of the resistors. 
With these approximations we get 
\begin{eqnarray}
&& iS_{\rm el}^\lambda = - \sum_{j=1,2} \int_0^t dt'dt''
\int\frac{d\omega}{2\pi} \frac{e^{-i\omega(t'-t'')}}{\omega R_j}
\nonumber\\ &&\times\,
\big[\eta_{j}^{++}(\omega) V_j^+(t') V_j^+(t'') + \eta_{j}^{+-}(\omega) V_j^-(t') V_j^+(t'')
\nonumber\\ &&
+\, \eta_{j}^{--}(\omega) V_j^-(t') V_j^-(t'') \big],
\label{action_res}
\end{eqnarray}
where the functions $\eta_{j}^{\alpha\beta}(\omega)$ are given by Eqs. (\ref{eta}) with photon distribution functions
averaged over the volume of the resistors, $n_j(\omega)=\int_{\Omega_j} d^3{\bm r}n_{\omega,\bm{r}}/\Omega_j$,
and with $\lambda_{\bm r}$ replaced by  $\lambda_j$.
The fields ${\bm E}$ and ${\bm H}$
in 3d space around the resistors and other circuit elements can be expressed via the voltages $V_j$ by solving
linear Maxwell equations with proper boundary conditions. In this way one finds
\begin{eqnarray}
{\bm E}_{F,B}(t,{\bm r})&=&\int_{-\infty}^t dt'\big[ {\bm e}_1(t-t',{\bm r})V_1^{F,B}(t')
\nonumber\\ &&
+\,{\bm e}_2(t-t',{\bm r})V_2^{F,B}(t')\big],
\label{E}
\\
{\bm H}_{F,B}(t,{\bm r})&=& \int_{-\infty}^t dt'\big[{\bm h}_1(t-t',{\bm r})V_1^{F,B}(t')
\nonumber\\ &&
+\,{\bm h}_2(t-t',{\bm r})V_2^{F,B}(t')\big],
\label{H}
\end{eqnarray}
where ${\bm e}_j(t,{\bm r})$ and ${\bm h}_j(t,{\bm r})$ are the fundamental solutions 
for electric and magnetic fields, which depend on the sample geometry. 
The solutions (\ref{E},\ref{H}) should be
substituted into the electro-magnetic part of the action (\ref{Sem}). After 
the integration over coordinates, this action becomes quadratic in the potentials $V_j$.
Moreover, since $E_F^2-E_B^2-H_F^2+H_B^2=2E^-E^+ - 2H^-H^+$ only the combinations $V_i^- V_j^+$ appear in it.
The coefficients in front of these combinations are expressed in terms of the functions ${\bm e}_j(t,{\bm r})$, ${\bm h}_j(t,{\bm r})$
and determine the impedances $Z_j(\omega)$, shown in Fig. 1a, for a given sample.  
Finally the electro-magnetic part of the action acquires the form 
\begin{eqnarray}
&& iS_{\rm em} =  -\int_0^t dt'dt''
\int\frac{d\omega}{2\pi}  
\frac{e^{-i\omega(t'-t'')}}{\omega} 
\nonumber\\&&\times\,
 \bigg[\sum_{j=1,2}\frac{V_j^-(t') V_j^+(t'')}{Z_j(\omega)}+\frac{V_{12}^-(t')V_{12}^+(t'')}{Z_0(\omega)}\bigg],
\label{action_em}
\end{eqnarray}
where $V_{12}^\pm=V_1^\pm-V_2^\pm$. 
According to our assumptions the impedances $Z_j(\omega)$ are purely imaginary, i.e. ${\rm Re}\,(1/Z_j)=0$. 
That is why the terms $\propto V^-(t')V^-(t'')$
do not appear in $iS_{\rm em}$. In contrast, such terms present in the action of the resistors (\ref{action_res}) even
if one puts $\lambda_1=\lambda_2=0$. These terms are related to dissipation in the resistors
and describe the current noise associated with it.  

At long observation time, $t\gg 1/T_j,1/\omega_c$,
the full action (\ref{S}) acquires the form
\begin{eqnarray}
iS^\lambda=\frac{i t}{2}\sum_n \vec V^T(-\omega_n) \frac{{\bm M}_\lambda(\omega_n)}{i\omega_n}\vec V(\omega_n),
\end{eqnarray}
where $\omega_n=2\pi n/t$ are discrete frequencies, 
$\vec V^T(\omega_n)=(V^+_1(\omega_n),V^-_1(\omega_n),V^+_2(\omega_n),V^-_2(\omega_n))$ is the vector of Fourier transformed voltages, 
and 
\begin{widetext}
\begin{eqnarray}
{\bm M}_\lambda(\omega_n)=\left(
\begin{array}{cccc}
-\frac{2\eta^{++}_1(\omega_n)}{R_1} & \frac{1}{Z_1^*}+\frac{1}{Z^*_0} - \frac{\eta_1^{+-}(-\omega_n)}{R_1} & 0 & -\frac{1}{Z^*_0}\\
-\frac{1}{Z_1}-\frac{1}{Z_0} - \frac{\eta_1^{+-}(\omega_n)}{R_1} & -\frac{2\eta^{--}_1(\omega_n)}{R_1} & \frac{1}{Z_0} & 0 \\
0 & -\frac{1}{Z^*_0} & -\frac{2\eta^{++}_2(\omega_n)}{R_2} & \frac{1}{Z_2^*}+\frac{1}{Z^*_0} - \frac{\eta_2^{+-}(-\omega_n)}{R_2}\\
\frac{1}{Z_0} & 0 & -\frac{1}{Z_2}-\frac{1}{Z_0} - \frac{\eta_1^{+-}(\omega_n)}{R_2} & -\frac{2\eta^{--}_1(\omega_n)}{R_2}
\end{array}
\right).
\end{eqnarray}
\end{widetext}
The Gaussian path integral (\ref{F2})
over $\vec V_n$ is evaluated exactly. Utilizing the property ${\bm M}_\lambda(\omega)=-{\bm M}_\lambda^T(-\omega)$ in the long time limit 
we find CGF in the form
$
F(t,\lambda) = -t\int_0^\infty \frac{d\omega}{2\pi}
\ln\big[{\det {\bm M}(\omega)}/{\det {\bm M}_{\lambda=0}(\omega)}\big]. 
$
Evaluating the determinants, and keeping in mind that $Z^*_j=-Z_j$ for reactive elements, we find
\begin{eqnarray}
&& F(t,\lambda) = -t\int_0^\infty \frac{d\omega}{2\pi} 
\ln\big[ 1-\tau(\omega)\big\{ n_1(\omega)[1+n_2(\omega)]
\nonumber\\ &&\times\,
\left(e^{i\lambda\omega}-1\right)
+[1+n_1(\omega)]n_2(\omega)\left(e^{-i\lambda\omega}-1\right)\big\}\big].\;\;\;\;
\label{FCS}
\end{eqnarray} 
Here $\lambda=\lambda_1-\lambda_2$,
\begin{eqnarray}
\tau(\omega)=\frac{4}{R_1R_2\left|G_1+G_2+Z_0G_1G_2\right|^2},
\label{transmission1}
\end{eqnarray}
is the effective transmission probability, and $G_j={1}/{R_j}+{1}/{Z_j(\omega)}$.

Equation (\ref{FCS}) is the main result of our paper.
It is the CGF of photons which are scattered
by a barrier with the transparency $\tau(\omega)$ and carry the energy $\omega$ each.
It is consistent with standard results of quantum optics \cite{Glauber}
and closely resembles the CGF of scattered electrons \cite{Levitov1}, which are fermions. 
In the context of photon scattering by a cavity similar expression has been derived by Beenakker \cite{Beenakker}, 
and in the context of phonon heat conductance --- by Saito and Dhar \cite{Saito1}.
If both $n_1(\omega)$ and $n_2(\omega)$ have the equilibrium Bose form, CGF (\ref{FCS}) acquires the 
property $F(\lambda)=F(-\lambda + i(T_1^{-1}-T_2^{-1}) )$, which translates into the fluctuation theorem
$P(Q)=P(-Q)\exp[Q(T_1^{-1}-T_2^{-1})]$. We remind that the Eq. (\ref{FCS}) has been derived assuming 
Gaussian fluctuations of currents and voltages in the electric circuit. That implies, in particular, 
that the resistors $R_1$ and $R_2$ are linear elements, which do not exhibit Coulomb blockade
or other types of non-linearities. Besides that we have assumed that the real parts of the impedances $Z_j(\omega)$ 
are equal to zero and they correspond to purely reactive elements like inductors, capacitors or
their arbitrary combinations.       

\begin{figure}
\includegraphics[width=8cm]{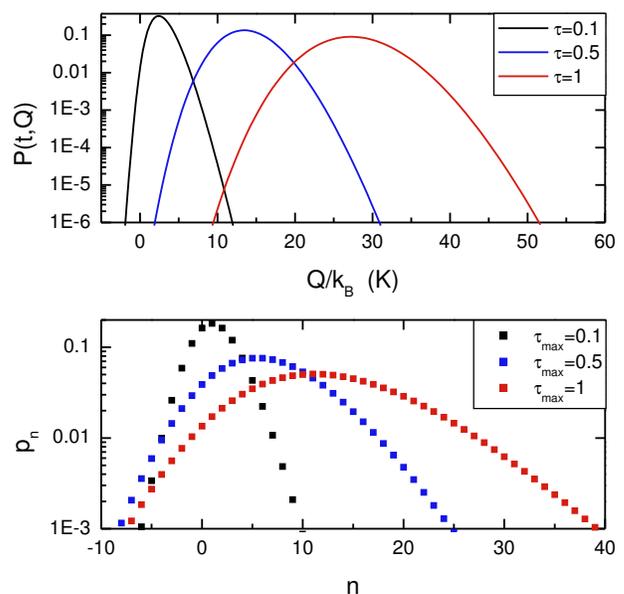}
\caption{Distribution of energy transmitted between the resistors during time $t$ for different 
transmission probabilities $\tau(\omega)$.  
(a) $\tau(\omega)=$const, $T_1=300$ mK, $T_2=100$ mK, the observation time is $t=10$ ns.
(b) $\tau(\omega)$ has the Lorentzian shape, $T_1=300$ mK, $T_2=100$ mK, $t=1$ ms,
CGF is given by Eq. (\ref{Fnarrow}). Discrete number of transferred photons $n$ is shown on the horizontal axis. }
\label{Fig_distribution}
\end{figure}

\section{Results and discussion}

Let us now consider some limiting cases. First we assume that the transmission probability, $\tau$,
is constant and the photon distribution functions have equilibrium Bose form. 
In this case the heat current acquires familiar form
$
J_Q=-(i/t)dF/d\lambda|_{\lambda=0}=\pi\tau\left(T_1^2-T_2^2\right)/12.
$
The simplest example of such a system is given by two directly connected resistors  (Fig. 1e), in which case
$\tau=4R_1R_2/(R_1+R_2)^2$. 
In Fig. \ref{Fig_distribution}a we show the distribution $P(t,Q)$ for three different values of $\tau$. 
The distribution becomes Gaussian at sufficiently long observation time such that $J_Qt\gg T_1$. 
The low frequency noise of the heat current is given by the expression
\begin{eqnarray}
&& S_Q=-\frac{1}{t}\frac{d^2F}{d\lambda^2}\bigg|_{\lambda=0}=
\left( \frac{\zeta(3)}{\pi}\tau(1-\tau) +\frac{\pi}{6}\tau^2 \right)(T_1^3+T_2^3)
\nonumber\\&&
+\, 2\tau(1-\tau)\int_0^\infty \frac{d\omega}{2\pi}\frac{\omega^2}{\left(e^{\omega/T_1}-1\right)\left(e^{\omega/T_2}-1\right)}.
\end{eqnarray}

Another interesting limit is transmission within a narrow Lorentzian 
with $\tau(\omega)=\tau_{\max}\Gamma^2/[(\omega-\omega_0)^2+\Gamma^2]$
and $\Gamma\ll\omega_0,T_1,T_2$. In this case
\begin{eqnarray}
F = -\Gamma t\left(\sqrt{1-\tau_{\max} f(\omega_0)}-1\right),
\label{Fnarrow}
\end{eqnarray}
where
$
f(\omega_0) = n_1(\omega_0)[1+n_2(\omega_0)]\left(e^{i\lambda\omega_0}-1\right) 
+[1+n_1(\omega_0)]n_2(\omega_0)\left(e^{-i\lambda\omega_0}-1\right).
$
Since $F(\lambda)$ becomes a periodic function of $\lambda$ in this approximation, we get
$
P(t,Q)=\sum_n p_n\delta(Q-n\omega_0)
$
with
$
p_n=\frac{\omega_0}{2\pi}\int_{-\pi/\omega_0}^{\pi/\omega_0} d\lambda \,e^{in\lambda\omega_0}e^{F(\lambda)}
$
being the probability to transmit $n$ photons with one frequency $\omega_0$. The distributions $p_n$
for three different values of $\tau_{\max}$ are shown in Fig. \ref{Fig_distribution}b.  
Due to the suppression of the average heat current between the resistors
the distributions $p_n$ significantly deviate from the Gaussian form even though the
observation time is long, $t=1$ ms. It is obvious from Eq. (\ref{Fnarrow}) that
the distribution $p_n$ becomes Poissonian in the limit $T_1\gg T_2$ and $\tau_{\max}\ll 1$.
At higher transparencies it deviates from the Poissonian form similarly to
what has been predicted in Ref. \onlinecite{Schomerus}, where the statistics of photons emitted
by a coherent conductor has been studied and rectangular shape of the transmission line has been assumed.
The average heat current and the noise corresponding to CGF (\ref{Fnarrow}) are (here $n_j\equiv n_j(\omega_0)$)
\begin{eqnarray}
&& J_Q=\frac{\tau_{\max}}{2}\Gamma\omega_0[n_1-n_2],\;\; S_Q = \frac{\tau_{\max}}{2}\Gamma\omega_0^2\big( n_1[1+n_2]
\nonumber\\ &&
+\,[1+n_1]n_2 + \tau_{\max} [n_1-n_2]^2/2\big).
\end{eqnarray}

\begin{figure}
\includegraphics[width=8cm]{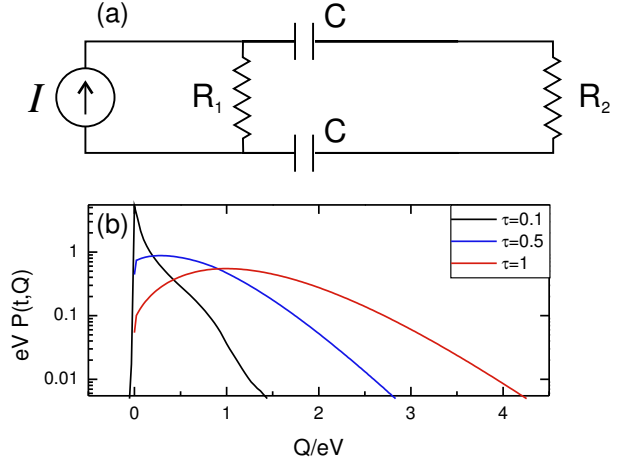}
\caption{ (a) Bias current $I$ is applied to the resistor 1 in order to drive it out of equilibrium. 
Two capacitors $C$, which shield the detector resistor 2 at low frequencies, 
are big enough to become fully transparent at frequencies
$\omega\sim\max\{T_1,T_2,eV\}$, where $V=IR_1$. In this case the barrier transmission 
$\tau$ may be approximately treated as frequency independent constant. 
(b) Distribution of transmitted energy during the observation time $t=100/eV$ for three
different values of  $\tau$. 
$Q$ and $P(Q)$ are scaled with the characteristic photon energy $eV$. }
\label{resistor-biased}
\end{figure}

Next we assume that leads are attached to the resistor 1 and bias current $I$ is applied to it (see Fig. \ref{resistor-biased}a). 
The electron distribution function inside it acquires a non-equilibrium double step form \cite{Nagaev},
$f(E,x)=(x/L_1)f(E-eV)+(1-x/L_1)f(E)$, where $V=IR_1$ is the voltage drop. 
We also assume that the temperatures of the resistor 2 and of the outer leads are much lower than $eV$. 
In this case one can put $n_2(\omega)=0$ and from the Eq. (\ref{n_photon}) we find $n_1(\omega)=(eV-\omega)\theta(eV-\omega)/6\omega$. 
Thus the CGF (\ref{FCS}) takes the form 
\begin{eqnarray}
F = -t\int_0^{eV} \frac{d\omega}{2\pi} \ln\left[ 1-\frac{\tau}{6}\frac{eV-\omega}{\omega} \left(e^{i\lambda\omega}-1\right)\right].
\label{FeV}
\end{eqnarray}
The corresponding distribution $P(Q)$ is shown in Fig. \ref{resistor-biased}b. It is strongly asymmetric with $P(Q)=0$ for $Q<0$,
i.e. over long intervals of time, $eVt\gtrsim 1$, the energy flows from the biased resistor
to the unbiased one, but never in the opposite direction. 
A somewhat similar system, namely a biased resistor coupled to an open transmission line, has been earlier
considered in Ref. \onlinecite{Ojanen}, where the average value of the heat current and its noise
have been derived. From CGF (\ref{FeV}) we find these parameters in our model
\begin{eqnarray}
J_Q=\frac{\tau(eV)^2}{24\pi},\;\;
S_Q=\left(1+\frac{\tau}{3}\right)\frac{\tau(eV)^3}{72\pi}.
\end{eqnarray}

\begin{figure}
\includegraphics[width=8cm]{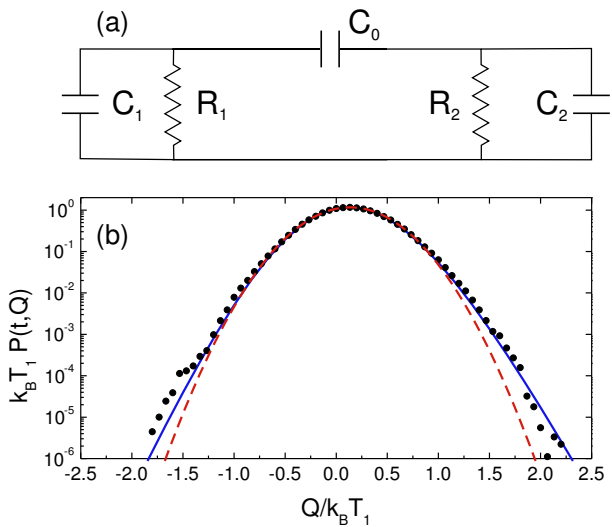}
\caption{ (a) Setup of the experiment \cite{Ciliberto1,Ciliberto2}.
The circuit parameters are: $R_1=R_2=10$ M$\Omega$, $C_0=100$ pF, $C_1=680$ pF, $C_2=420$ pF. 
The parameters defined in the text take the values $\alpha=2.134$, $\beta=0.0506$, and $t_0=6.29$ ms. 
(b) Distribution of energy transmitted during the time $t=0.1$ sec and for
resistor temperatures $T_1=296$ K, $T_2=88$ K.
Circles -- experimental points\cite{Ciliberto1,Ciliberto2}; 
blue line -- Eq. (\ref{Pclassical}); red dashed line -- Gaussian 
approximation $P(t,Q)=\exp[-(Q-J_Qt)^2/2S_Qt]/\sqrt{2\pi S_Qt}$, where $J_Q$ and $S_Q$ are defined
by Eqs. (\ref{SPcl}).}
\label{Ciliberto_fig}
\end{figure}

In the classical limit $T_j\gg\omega_c$ CGF (\ref{FCS}) reduces to  \cite{Saito2}
\begin{eqnarray}
F=-t\int_0^\infty \frac{d\omega}{2\pi}\ln\left[1-\tau(\omega)\left( i\lambda\Delta T_{12}-\lambda^2T_1T_2 \right)\right],
\label{Fclassical}
\end{eqnarray}
where $\Delta T_{12}=T_1-T_2$.
It is interesting to compare this result with the experiment\cite{Ciliberto1,Ciliberto2}. 
In that experiment capacitors have been used, which implies $Z_j(\omega)=1/(-i\omega C_j)$ (see Fig. \ref{Ciliberto_fig}a).  
Accordingly, $\tau(\omega)$  (\ref{transmission1}) takes the form 
\begin{eqnarray}
\tau(\omega)=\frac{2\beta (\omega t_0)^2}{1+2(\alpha-1) (\omega t_0)^2 + (\omega t_0)^4},
\label{transmission2}
\end{eqnarray}
with
$
t_0=\sqrt{R_1R_2(C_1C_2+C_0C_1+C_0C_2)}
$,
\begin{eqnarray}
\alpha=1+[R_1^2(C_0+C_1)^2+R_2^2(C_0+C_2)^2+2R_1R_2C_0^2]/2t_0^2,
\nonumber
\end{eqnarray}
and $\beta={2C_0^2}/(C_1C_2+C_0C_1+C_0C_2)$.
For this model one can exactly evaluate CGF (\ref{Fclassical}),
\begin{eqnarray}
F=\frac{t}{t_0}\left(\sqrt{\frac{\alpha}{2}}-\sqrt{\frac{\alpha}{2}-\frac{\beta(i\lambda\Delta T_{12}-\lambda^2T_1T_2)}{2}}\right),
\end{eqnarray}
and the distribution of the transferred heat $P(t,Q)=\int\frac{d\lambda}{2\pi}e^{-i\lambda Q + F(t,\lambda)}$, which reads
\begin{eqnarray}
P(t,Q) &=& \frac{t}{\pi}\sqrt{\frac{2a}{\beta T_1T_2t^2+2Q^2t_0^2}}\,
e^{\frac{t}{t_0}\sqrt{\frac{\alpha}{2}}+\frac{T_1-T_2}{2T_1T_2}Q}
\nonumber\\ &&
K_1\left(\sqrt{a\left(\frac{t^2}{t_0^2}+\frac{2Q^2}{\beta T_1T_2}\right)}\right).
\label{Pclassical}
\end{eqnarray}
Here $K_1(x)$ is the modified Bessel function of the second kind, and
$
a=\alpha/2+\beta(T_1-T_2)^2/8T_1T_2.
$
One should bear in mind that the expression (\ref{Pclassical}) is valid
in the long time limit $t\gtrsim t_0$.  
The average heat current from the resistor 1 to the resistor 2 and the corresponding noise in this model have the form
\begin{eqnarray}
J_Q=\frac{\beta(T_1-T_2)}{2\sqrt{2\alpha} t_0},\;\;
S_Q = \frac{\beta T_1T_2}{\sqrt{2\alpha}\, t_0} + \frac{\beta^2(T_1-T_2)^2}{4\sqrt{2}\alpha^{3/2} t_0}.
\label{SPcl}
\end{eqnarray}
We compare the distribution (\ref{Pclassical}) with the experimental one \cite{Ciliberto1,Ciliberto2}
in Fig. \ref{Ciliberto_fig}b. The agreement between the two is quite good. 
In particular, one can see the deviations from Gaussian form at the tails of the distribution.
The subtle point of the measurements \cite{Ciliberto1,Ciliberto2} was the difference between the heat $Q_1$, i.e.
the change of the energy of the resistor 1, and the work $W_1$, which also includes the change of the electrostatic
energy of the capacitor $C_1$. We have verified that in the long time limit  both $Q_1$ and $W_1$ should
have the same distribution (\ref{Pclassical}). On the qualitative level this can be understood 
from the relation $W_1=Q_1 + C_1[V_1^2(t)-V_1^2(0)]/2$. Indeed, the average value of the last term, i.e. of the change in the 
energy of the capacitor $C_1$ during the observation time $t$, equals to zero because
$\langle V_1^2(t)\rangle$ is finite and does not grow in time.  
Since both $Q_1$ and $W_1$ grow in time linearly, one can put $W_1\approx Q_1$ at sufficiently long $t$
even without averaging. Experimentally, however, the work distribution has approached
the long time limit form faster than the heat distribution.  
That is why in Fig. \ref{Ciliberto_fig}b we plot the experimental work distribution $P(W_1)$.
Further analysis is required in order to understand the origin of this behavior.

We propose the distribution of heat in the low temperature quantum regime to be measured in the setup similar
to the one used in the experiments [\onlinecite{Meshke},\onlinecite{Timofeev}]. Namely, 
one would monitor the temperature of the detector resistor 2 in real time
with the time resolution of the order of $t_0\approx 12\hbar/\pi\tau k_BT_1$,
that is the time interval during which an average energy $k_BT_1$ is transferred from the resistor 1 to the resistor 2. 
Assuming $T_1=100$ mK and $\tau=3\times 10^{-4}$ one finds $t_0\approx 1$ $\mu$s, which
is within the reach of current technology\cite{Simone}. The expected magnitude
of temperature fluctuations in the second resistor caused by fluctuations
of heat flow may be estimated as 
$\delta T_2^2 \approx 3\tau T_2 t/\pi^3\hbar k_B\nu^2\Omega_2^2$, where $t$ 
is the observation time. For a resistor with the volume $\Omega_2=0.001$ $\mu$m$^3$ made of copper
(density of states $\nu\approx 10^{29}$ J$^{-1}$ $\mu$m$^{-3}$) and for $T_2=50$ mK and $t=100 t_0$
one finds  $\delta T_2\sim 15$ mK, which is measurable with currently available
thermometers based on normal metal -- superconductor tunnel junctions\cite{Simone,Klara}.  
One can further optimize the system by, for example, 
designing the coupling circuit with narrow line transmission spectrum, or by
using other types of temperature sensors like, e.g., recently proposed sensor based on an SNS Josephson
junction \cite{Tero,Jonas}.  

In summary, we have developed a theory of full counting
statistics of heat exchange between two metallic resistors, which is valid both at high and at low temperatures, 
where the classical formula for the noise (\ref{Nyquist}) can no longer be used.
Fluctuations of the heat current in this system
can be interpreted as scattering of photons by an effective potential barrier.
In high temperature limit our results are in good agreement with recent experiment\cite{Ciliberto1,Ciliberto2}. 
We acknowledge very useful discussions with S. Ciliberto, G. Lesovik, O. Saira and Y. Utsumi.
We are grateful to S. Ciliberto for providing us with the experimental data.
This work has been supported in
part by the Academy of Finland (projects no.
272218 and 284594), and the European Union Seventh
Framework Programme INFERNOS (FP7/2007-
2013) under grant agreement no. 308850.

\end{document}